\documentclass[aps,amssymb,nofootinbib]{revtex4}

\usepackage{amssymb,amsmath}
\usepackage{amsfonts}
\usepackage{mathrsfs}
\usepackage{txfonts}
\usepackage{marvosym}

\begin{document}


\title{A unique Fock quantization for fields in
non-stationary spacetimes}

\author{Jer\'onimo Cortez} \email{jacq@fciencias.unam.mx}
\affiliation{Departamento de F\'\i sica,
Facultad de Ciencias, Universidad Nacional Aut\'onoma de
M\'exico,
 M\'exico D.F. 04510, Mexico.}
\author{Guillermo A. Mena
Marug\'an}\email{mena@iem.cfmac.csic.es}
\affiliation{Instituto de Estructura de la Materia, CSIC,
Serrano
121, 28006 Madrid, Spain.}
\author{Javier Olmedo}\email{olmedo@iem.cfmac.csic.es}
\affiliation{Instituto de Estructura de la Materia,
CSIC, Serrano 121, 28006 Madrid, Spain.}
\author{Jos\'e M. Velhinho}\email{jvelhi@ubi.pt}
\affiliation{Departamento de F\'{\i}sica, Universidade
da Beira Interior, R. Marqu\^es D'\'Avila e Bolama,
6201-001 Covilh\~a, Portugal.}

\begin{abstract}
In curved spacetimes, the lack of criteria for the
construction of a unique quantization is a fundamental
problem undermining the significance of the
predictions of quantum field theory. Inequivalent
quantizations lead to different physics. Recently,
however, some uniqueness results have been obtained
for fields in non-stationary settings. In particular,
for vacua that are invariant under the background
symmetries, a unitary implementation of the classical
evolution suffices to pick up a unique Fock
quantization in the case of Klein-Gordon fields with
time-dependent mass, propagating in a static spacetime
whose spatial sections are three-spheres. In fact, the
field equation can be reinterpreted as describing the
propagation in a Friedmann-Robertson-Walker
spacetime after a suitable scaling of the field by a
function of time. For this class of fields, we prove
here an even stronger result about the Fock
quantization: the uniqueness persists when one allows
for linear time-dependent transformations of the
field in order to account for a scaling by background
functions. In total, paying attention to the dynamics,
there exists a preferred choice
of quantum field, and only one $SO(4)$-invariant Fock
representation for it that respects the standard
probabilistic interpretation along the evolution. The
result has relevant implications e.g. in cosmology.
\end{abstract}


\maketitle

\section{Introduction}
In inhomogeneous cosmology, the study of quantum phenomena
is often affected by two related types of ambiguities which
influence the physical predictions, and which
appear owing to the infinite number of degrees of freedom that are
present in the inhomogeneities. The first of these
ambiguities concerns the choice of variables (usually scalar
or tensor fields) which are employed to describe, or
{\emph{parameterize}}, the local degrees of freedom of the system.
The second ambiguity is generic in quantum field theory, and
concerns the selection of an appropriate quantum representation
for the field theory obtained once a choice of field variables is
made.

The freedom in the choice of fields has in fact
potentially non-trivial consequences. Namely, in non-stationary
and spatially homogeneous scenarios, like those provided by the
typical kind of background solutions encountered in cosmology,
different choices of ``fundamental'' fields arise naturally from
time-dependent transformations, each of them leading in general to
different dynamics. Normally, these transformations
consist in a scaling of the fields by time-dependent functions, so that
the linearity of the field equations and of the structures of the system are
maintained. But, while in theories with a finite number of degrees of freedom
these linear transformations can be promoted generally to quantum unitary
operators, the situation changes drastically
in quantum field theory. In this latter case, linear transformations do not always
admit a unitary implementation and distinct dynamics
usually call for inequivalent representations. Therefore, it is clear
that the final quantum theory depends intimately on which fields
(and hence on which dynamics) are viewed as the fundamental ones.

An important class of systems for which this kind of considerations
has a major relevance are quantum matter fields propagating in a
classical non-stationary spacetime, like e.g. the case of matter fields
in inflationary backgrounds. Another large class is found in the
quantization of local gravitational degrees of freedom (in contraposition
to the quantization of matter in gravitational backgrounds). This includes
inhomogeneous models with certain types of symmetries, such as the
so-called Gowdy models \cite{gowdy}, as well as the quantization of
gravitational perturbations around homogeneous, cosmological backgrounds.
For this latter class, the parametrization of the metric components by
means of appropriate fields is only constrained by symmetry arguments,
usually leaving an ample freedom in the choice that can be reinterpreted
in terms of a time-dependent scaling of the fields. In addition, in the case
of matter fields, it is often convenient to scale them by a suitable
combination of background functions, typically related
to the scale factor of the background spacetime, but whose specific form
depends on the particular system under discussion.\footnote{As an example, let us
mention the case of the gauge-invariant energy density per\-turbation
amplitude in perfect fluid cosmologies; see e.g. \cite{bardeen}.}

For this kind of non-stationary settings, the commented freedom in the
choice of fields often allows one to select --as fundamental ones--
fields which effectively live in an auxiliary static background, although
generally subject to time-dependent potentials. This permits one to simplify
the corresponding dynamics at least partially. For instance,
let us mention again the case
of the Gowdy cosmologies \cite{gowdy}, which are spacetimes with two
spacelike Killing isometries and a compact spatial topology.
In the particular case of a three-torus topology and a content of linearly
polarized gravitational waves, the local gravitational degrees of freedom
of these cosmologies can be described by a single scalar field defined on
the circle. The field evolves with respect to a quadratic and explicitly
time-dependent Hamiltonian which, for a particular choice of field
parametrization, can be seen as the Hamiltonian of a free particle on the
circle with a time-dependent mass \cite{unit-gt3}. Similar descriptions can
be obtained for the Gowdy models with the other possible spatial topologies,
namely the three-sphere and the three-handle \cite{gowdy}, with the
difference that in those cases the field is effectively defined on the
sphere, $S^2$.

Also in the case of (inhomogeneous) cosmological
perturbations, and more straightforwardly for quantum test fields in a
non-stationary gravitational background, one can map the theory to a field
model in a static background. For instance, for scalar matter fields
coupled to Friedmann-Robertson-Walker (FRW) spacetimes (with applications
e.g. to inflation \cite{mukhanov,varios,lyth}), a Klein-Gordon equation
for the matter field becomes, after a re-scaling in terms of the scale factor,
a linear wave equation with a time-dependent mass, but now in a
static auxiliary background with the same spatial topology and dimension
as the FRW cosmology. On general grounds, the dynamics of a massive scalar
field $\varphi$ in an FRW spacetime $ds^2=a^{2}(t)[-dt^2+\gamma_{ij}dx^i dx^j]$
--where $\gamma_{ij}$ ($i=1,2,3$)
is the standard Riemannian metric of either a three-sphere,
a three-dimensional flat space, or a three-dimensional hyperboloid--
is governed by the equation
\begin{equation}\ddot{\varphi}+2\frac{\dot{a}}{a}\dot{\varphi}-
\Delta\varphi +m^{2}a^{2}\varphi =0.
\end{equation}
The dot denotes time derivative with respect to the conformal time $t$,
and $\Delta$ is the Laplace-Beltrami operator associated with the spatial metric
$\gamma_{ij}$. It is then straightforward to check that, if one introduces the
time-dependent scaling $\phi=a\varphi$, the dynamics in the new field description
is dictated by
\begin{equation}
\ddot{\phi}-\Delta\phi+s(t)\phi =0,
\end{equation}
where $s(t)=m^{2}a^{2}-(\ddot{a}/a)$. The system can now be
treated as a Klein-Gordon field propagating in a static background
$ds^2=-dt^2+\gamma_{ij}dx^idx^j$ but in the presence of a time-varying potential
$V(\phi)=s(t)\phi^{2}/2$. Notice that $s(t)$ can be interpreted as a
nonnegative time-dependent mass $m(t)=s^{1/2}(t)$, provided that $s(t)\geq 0$. The same
arguments can be applied when the scalar field coupled to the FRW spacetime is also
subject to a time-dependent potential quadratic in the field.

On the other hand, as we have remarked, even if the field parametrization of the
system is specified, there exists a second type of ambiguity that affects the
physical predictions. This concerns the selection of a quantum
representation for the field theory in hand: in the case of fields,
with local degrees of freedom, the system admits infinite non-equivalent
representations of the canonical commutation relations (CCR's) \cite{wald}
and there is no general procedure to select a preferred quantum description.

The usual strategy to pick up a distinguished representation in a
given field theory is to exploit the classical symmetries. For instance,
invariance under the Poincar\'e group, adapted to the theory under consideration,
is the criterion imposed to arrive at a unique representation in ordinary
quantum field theory.
In particular, for {\emph{a}} scalar field theory, this imposition of
Poincar\'e invariance selects a complex structure \cite{wald},
which is the mathematical object that encodes the ambiguity in the quantization
and determines the vacuum state of the Fock representation.\footnote{Poincar\'e
invariance selects in fact a continuous family of representations,
characterized by a mass parameter. A unique element of that family
is obtained when the Poincar\'e group is properly adapted to the dynamics of the system,
since one of the generators of the group is a time-like Killing vector which provides a
natural decomposition in positive and negative frequencies.}
For systems with time translation invariance,
this symmetry is also exploited in order to  formulate the so-called energy
criterion and then choose a preferred complex structure \cite{a-m}. But
when the symmetries are  restricted, as it is the case for non-stationary
spacetimes or for manifestly time-dependent systems, extra requirements
must be imposed to complete the quantization process. E.g., for
a free scalar field in $1+1$ de Sitter spacetime, a unique de Sitter
invariant Fock vacuum is selected by looking for an
invariant Gaussian solution to a regulated Schr\"{o}dinger equation
\cite{jackiw}.

Recently, the unitary implementation of the dynamics has been
successfully employed as an additional criterion to specify a
unique Fock quantization for the linearly polarized Gowdy model
with three-torus topology. In particular, it has been shown
that there is essentially one field parametrization and one
Fock representation which is invariant under the symmetries of
the (auxiliary) background and allows a unitary implementation
of the corresponding field dynamics \cite{unique-gowdy-1,CMV5}. Thus, both
ambiguities have been resolved in that case. Furthermore, by demanding
a unitary dynamics and invariance under the classical symmetries, unique
Fock representations have been specified as well for free scalar fields
with generic time-dependent mass terms defined on the circle, the two-sphere,
or the three-sphere \cite{CMV8}, thus endorsing the unitarity criterion.
As we have pointed out, the case of the two-sphere covers the other two
possible topologies of the Gowdy cosmologies. On the other hand, a scalar
field with time-dependent mass on the three-sphere not only describes test
fields in an FRW spacetime, but in addition finds applications in the
treatment of cosmological perturbations \cite{mukhanov,mfb}. In this
perturbative framework realistic scenarios are explored by considering
small inhomogeneous departures from FRW spacetimes, and one frequently
has to deal with linear wave equations with a quadratic time-dependent
potential (for a brief account, see e.g. \cite{CMV8}). Thus, as far as
this entire class of cosmological systems is concerned,
and restricting the discussion to the {\em specific} field parametrization
considered, our previous results assert that there is a unique Fock
representation such that the vacuum state has the symmetries of the
background and the evolution is unitarily implementable.

In this article we go beyond those results by analyzing the remaining
freedom in the choice of fields. As we have already mentioned,
the freedom that we are interested in considering amounts to
time-dependent scalings of the field, which respect the linearity of
the field equations and hence of the space of solutions. We will therefore study
the most general compatible linear canonical transformation, which
consists of a time-dependent scaling of the configuration variable, the
inverse scaling of its canonical momentum, and the possible addition to
this momentum of a time-dependent contribution which is also
linear in the configuration variable.
We will show that, unless the transformation is trivial,
the dynamics of the transformed field is such that it is impossible
to find a Fock representation which is invariant under the background
symmetries and allows a unitary implementation of the evolution.
In this respect, the inclusion in the transformed momentum of a
contribution linear in the configuration
variable is important because, in practice, one usually starts with a
field description other than the one with the privileged dynamics.
Then, in general, the latter can be reached only if such a contribution is
taken into account.\footnote{Related to this fact,
this kind of contribution is necessary
if one wants to consider time-dependent transformations between field
descriptions whose quadratic Hamiltonian is free of
undesirable terms containing the product of the configuration
and momentum field variables (see e.g. the discussion in \cite{CMV5}).}

\section{Classical considerations}

Before we turn to the proof of our statement about the uniqueness
of the choice of field under time-dependent scalings in the case of
the three-sphere, let us describe the classical set-up of our theory.

We consider a real scalar field on $S^3$, with field equation
\begin{equation}
\label{sn-fieldequation}
\displaystyle
\ddot{\phi}-\Delta\phi + s(t)\phi =0 \, ,
\end{equation}
where the dot stands for the time derivative, $\Delta$
is now the Laplace-Beltrami operator corresponding to the
standard metric on $S^3$ and $s(t)$ is an arbitrary
time function (apart from some extremely mild
conditions about its derivative, stated in
\cite{CMV8}). After a decomposition of the field in
terms of (hyper)spherical harmonics (see e.g. \cite{har}), the
degrees of freedom are encoded in a discrete set of
modes $q_{n\ell m}$, where $n$ is any non-negative
integer, and the other two integers $\ell$ and $m$
vary from zero to $n$ and from $-\ell$ to $\ell$,
respectively. The modes obey decoupled equations of
motion
\begin{equation} \label{q-eq}
\displaystyle
\ddot q_{n\ell m}+[\omega_n^2+s(t)]q_{n\ell m}=0 \, ,
\end{equation}
where $\omega_n^2=n(n+2)$ are the eigenvalues of the
Laplace-Beltrami operator. These equations are
independent of the labels $\ell$ and $m$, and
therefore, for each $n$, there is a degeneracy of
$g_n=(n+1)^2$ modes with the same dynamics.

We restrict our analysis to the sector of non-zero
modes from now on. Obviously, the inclusion or not of
the zero mode does not affect the properties related to
the presence of an infinite number
of degrees of freedom. Let us then introduce the
annihilation-like variables
\begin{equation}
\label{basic-var}
\displaystyle
a_{n\ell m}={\frac{1}{\sqrt{2\omega_n}}} \left(\omega_nq_{n\ell
m} +i p_{n\ell m}\right)
\end{equation}
which, together with their complex conjugates (the
creation-like variables $a_{n\ell
m}^*$), provide a complete set of kinematical variables, or coordinates,
for the
phase space of the considered sector. Here, $p_{n\ell
m}={\dot q}_{n\ell m}$ is the canonically conjugate momentum of the
configuration variable $q_{n\ell m}$. The evolution of these
variables can be expressed as a linear transformation
which is block-diagonal (i.e., respects the harmonic
labels) and has the form
\begin{equation}
\label{bogo-transf}
\displaystyle
a_ {n\ell m}(t)=\alpha_{n}(t)a_
{n\ell m}(t_0)+ \beta_{n}(t)a_{n\ell m}^*(t_0) \, .
\end{equation}
The time functions $\alpha_n(t)$ and $\beta_n(t)$,
which depend as well on the reference time $t_0$,
completely characterize the classical evolution
operator (see \cite{CMV8} for details).

According to our previous discussion, we now consider
a time-dependent canonical transformation of the form
\begin{equation}
\label{transform}
\displaystyle
 \varphi:=f(t) \phi, \qquad
P_{\varphi}:=\frac{P_{\phi}}{f(t)}+g(t)\sqrt{h}\phi \, ,
\end{equation}
where $h$ is the determinant of the metric on $S^3$.
Our aim is to investigate the unitary implementability
of the dynamics of the new canonical fields
$(\varphi,P_{\varphi})$ obtained with this
transformation. It is shown in \cite{CMV5} that there
is no loss of generality in arbitrarily fixing the
transformation at the reference time $t_0$ so that
$f(t_0)=1$ and $g(t_0)=0$, conditions that we assume
in the following. Furthermore, we require that the
functions $f$ and $g$ be real and differentiable, and
that $f(t)$ vanish nowhere, so that the transformation
does not spoil the differential formulation of the
field theory, nor introduces singularities.

The dynamics of the transformed fields admits a
description like (\ref{bogo-transf}), but with new
coefficients $\tilde \alpha_n(t)$ and $\tilde
\beta_n(t)$, related to the old ones by:
\begin{eqnarray}
\label{26}
\displaystyle
\tilde \alpha_n(t)   &=& f_+(t)\alpha_n(t) +
f_-(t)\beta^*_n(t) +  \frac {i}{2}
\frac{g(t)}{\omega_n}[\alpha_n(t)+\beta_n^*(t)] \, ,
\\
\label{27}
\tilde \beta_n(t)   &=&  f_+(t) \beta_n(t) +
f_-(t)\alpha^*_n(t)+\frac{i}{2}\frac{g(t)}{\omega_n}[
\alpha^*_n(t)+\beta_n(t)] \, ,
\end{eqnarray}
where $2 f_{\pm}(t):=f(t)\pm 1/f(t)$.

\section{Uniqueness of the quantization}
\subsection{Symmetries and unitarity condition}

The possible Fock quantizations of the system are effectively determined
by the different complex structures that can be defined on phase space.
Strictly speaking, a complex structure $J$ is
a real linear transformation on phase space
which is compatible with its canonical symplectic structure,\footnote{Namely,
$J$ must be a symplectomorphism and, if $\Omega(\cdot,\cdot)$ is the
symplectic form, the bilinear map $\Omega(J\cdot,\cdot)$ must be positive definite.
This requirement guarantees that
$[\Omega(J\cdot,\cdot)- i \Omega(\cdot,\cdot)]/2$
provides an inner product.} and whose square is
minus the identity, $J^2=-{\bf 1}$.
Together with the symplectic structure, a complex structure
defines a state, usually called the vacuum, and hence a representation
of the CCR's.\footnote{To be precise, a complex structure
determines a state of the Weyl algebra generated by
the configuration and momentum field variables, from which a representation
of the CCR's can be defined (see e.g. \cite{CC} for a general introduction).}
The corresponding positive and negative frequency components are
obtained, respectively, with the projections $({\bf 1}-iJ)/2$ and
$({\bf 1}+iJ)/2$.

For the systems under consideration, as we have seen,
one can coordinatize the phase space by the
creation and annihilation-like variables (\ref{basic-var}), associated with the
harmonic modes. Using these variables, it was shown in \cite{CMV8} that the complex
structures that remain invariant under the group $SO(4)$ of rotations on the
three-sphere can be parameterized by a sequence of complex pairs
$(\kappa_n,\lambda_n)$, with
$|\kappa_n|^2-|\lambda_n|^2=1$ for every integer $n>0$.
Given the relation between Fock quantizations and complex structures,
it follows that the quantizations whose vacuum state is invariant under
the symmetry of the field equations, $SO(4)$, are again characterized by the
sequence of  pairs $(\kappa_n,\lambda_n)$ \cite{CMV8}.

Among the set of those $SO(4)$-invariant Fock representations, there still exist
infinitely many non-equivalent ones. We then impose the unitary
implementation of the dynamics as an additional criterion to select
a Fock representation among them. Unitary implementation of a given canonical
transformation essentially means that one is able to define in a consistent way
a quantum version of the transformation by means of the action of a unitary operator.
In the present case, we require this for all the canonical transformations
generated by the dynamics, i.e. for the entire set of transformations (\ref{bogo-transf})
that describe the evolution to all possible final times $t$.\footnote{We emphasize
that the functions appearing in (\ref{bogo-transf}) depend
on the field parametrization. After the transformation (\ref{transform}),
the general form of those functions is given in (\ref{26}) and
(\ref{27}).}
Let us then suppose that we have a certain Fock quantization, i.e. we are given
a Hilbert space and operators $\hat a_{n\ell m}$ and $\hat a^{\dagger}_{n\ell m}$
corresponding to our classical observables. When the transformation
(\ref{bogo-transf}) is applied to our quantum operators, we obtain new operators:
\begin{equation}
\label{quantumbogo}
\hat a_{n\ell m}(t):=\alpha_n(t)\hat a_{n\ell m}+\beta_n(t)\hat a^{\dagger}_{n\ell m}.
\end{equation}
By construction, the set of operators  $\hat a_{n\ell m}(t)$, together with their
adjoints, gives us a new representation of the CCR's.
We say that the classical transformation
(\ref{bogo-transf}) admits a unitary implementation when
this new representation is unitarily equivalent to the original one.
If this is the case, there indeed exist unitary operators $\hat{\cal U}(t)$
corresponding to the classical transformation, i.e., such that
$\hat a_{n\ell m}(t)=\hat{\cal U}^{-1}(t)\hat a_{n\ell m}
\hat{\cal U}(t)$.
In standard quantum mechanics the condition for unitary implementation is
trivially satisfied but, as mentioned in the Introduction, this is not generally so
when an infinite number of degrees of freedom are present.

As far as Fock quantizations are concerned, the unitary implementation of a linear
canonical transformation essentially depends on the effect of the transformation on
the vacuum of the considered representation. In fact, each Fock representation
comes with its own set of creation and annihilation operators, and the application
of a linear canonical transformation leads to new ones, like e.g. in
(\ref{quantumbogo}). One can then ask
whether the formal state in the kernel of all the new annihilation operators,
interpretable as a new vacuum, is a well-defined element of the original Hilbert
space. If the answer is in the affirmative, the transformation is unitarily
implementable.\footnote{In rigorous terms,
a canonical transformation $T$ is unitarily  implementable in a Fock
representation defined by a complex structure $J$ if and only if the operator
$T+JTJ$ is of the Hilbert-Schmidt type, on the one-particle Hilbert space defined by
$J$ \cite{hr-sh}.} Otherwise, it maps the original Fock
representation into a new, non-equivalent one. Actually, it is possible to see that
the new vacuum belongs to the original
Hilbert space if and only if its relation with the initial vacuum involves a finite
particle production. For a linear transformation of the
form (\ref{quantumbogo}) and assuming that the original vacuum is annihilated by all the
operators $\hat{a}_{nlm}$, for instance, the condition for a unitary implementation becomes
then just the summability over all modes ($n$, $l$, and $m$) of the square complex norm
of the functions $\beta_n(t)$ corresponding to this transformation.

It is worth noticing that the criterion that we are using here is just the
unitary implementation of the transformations which correspond to the evolution of the
system in \emph{finite} intervals of time, in opposition to the technically
more involved condition of the existence of a well-defined self-adjoint quantum
Hamiltonian (which in the cases under study is necessarily time dependent).
For a discussion  on the existence of a quantum
Hamiltonian for linear theories see e.g. \cite{helf}.
On a different context, the review article \cite{DE} also provides a modern
approach to unitarity issues in field theory, starting from the time-dependent
quantum harmonic oscillator.

\subsection{Uniqueness result}

We are now in adequate conditions to show that, after applying a
transformation of the considered type
(\ref{transform}), the dynamics becomes such that one cannot attain
a unitary implementation of the corresponding evolution
with respect to any $SO(4)$-invariant Fock representation.

Let us first make the unitary implementability condition fully explicit in our case.
Suppose that we are given  a $SO(4)$-invariant Fock representation of the CCR's,
determined by a sequence of  pairs
$(\kappa_n,\lambda_n)$ as explained above.
It is shown in detail in \cite{CMV8} (see also \cite{CMV5})
that the dynamics associated with the transformed canonical pair
$(\varphi,P_{\varphi})$ is unitarily implementable in
the considered $SO(4)$-invariant Fock quantization if and
only if the sequences
$\{\sqrt{g_n}\tilde\beta^J_n(t)\}$ are square summable
over $n$ for all possible values of $t$, where
\begin{equation}
\displaystyle
\tilde\beta^J_n(t):=(\kappa_n^*)^2\tilde\beta_n(t)-\lambda_n
^2\tilde\beta_n^*(t)+2i\kappa_n^*\lambda_n
\mathop{\rm Im}[\tilde\alpha_n(t)] \, .
\end{equation}
We will now see  that this summability condition can
only be fulfilled if the transformation
(\ref{transform}) is in fact the identity
transformation, i.e., if $f(t)$ is the identity
function and $g(t)$ vanishes. The arguments, whose
complete technical details  will appear in
\cite{CMOV}, are analogous to those presented in
\cite{CMV5}.

So, let us assume that
$\{\sqrt{g_n}\tilde\beta^J_n(t)\}$ is square summable
at all times. This implies that, for every $t$, the
terms of this sequence must tend to zero in the limit
of infinite $n$. Then, the same must occur with
$\tilde\beta^J_n(t)/(\kappa_n^*)^2$, since both $g_n$
and $|\kappa_n|$ are larger than 1. By substituting in
the expression of $\tilde\beta^J_n(t)$ the asymptotic
limits of $\alpha_n(t)$ and $\beta_n(t)$ (which were
analyzed in \cite{CMV8}), as well as that of
$\omega_n$, we conclude that the sequences given by
\begin{equation}
\label{z1}
\displaystyle
\left[e^{i(n+1)\tau}-\frac{\lambda_n^2}{(\kappa_n^*)^2}
e^{-i(n+1)\tau}\right]\frac{f_-(t)}{2}
-i\frac{\lambda_n}{
\kappa_n^*}\sin\left[(n+1)\tau\right]f_+(t)
\end{equation}
tend to zero for all values of $t$ when $n\to\infty$.
Here, $\tau:=t-t_0$.

We are now in a situation completely similar to that
studied in \cite{CMV5}. Applying the type of arguments
presented in Appendix A of that reference one can
prove that, if (\ref{z1}) tends indeed to zero at all
times, and hence the same happens with its imaginary
part, then the sequences with terms
$1-\mathop{\rm Re}\left[{\lambda_n^2}/{(\kappa^*_n)^2}\right]$ and
$\mathop{\rm Im}\left[{\lambda_n^2}/{(\kappa^*_n)^2}\right]$
cannot tend simultaneously to zero on any (infinite)
subsequence of the natural numbers.

Let us now restrict our attention to expression
(\ref{z1}) for the particular set of values $\tau=2\pi
q/p$, where $q$ and $p$ are arbitrary positive
integers subject only to the condition that
$t=\tau+t_0$ belongs to the allowed domain for this
time parameter. For each fixed value of $p$, we then
consider the subsequence of natural numbers of the
form $n=mp-1$, where $m$ can take any positive integer
value. Since the real and imaginary parts of
(\ref{z1}) must tend to zero on all subsequences, it
follows that both
$\left(1-\mathop{\rm Re}\left[{\lambda_{mp-1}^2}/{(\kappa^*_{mp-1})^2}
\right] \right)f_-\left(t_0+{2\pi q}/{p}\right) $ and
$ \mathop{\rm Im}\left[{\lambda_{mp-1}^2}/{(\kappa^*_{mp-1})^2}\right]
f_-\left(t_0+{2\pi q}/{p}\right) $ must approach a
vanishing limit as $m$ goes to infinity, for every
possible value of $p$ and $q$. However, we know that
the time-independent coefficients in these expressions
cannot tend simultaneously to zero on any subsequence
of the natural numbers. Therefore, our conditions can
only be fulfilled if $f_{-}(t_0+2\pi q/p)$ vanishes
for  all the possible values of $p$ and $q$ or,
equivalently, if
\begin{equation}
\displaystyle
f^2\left(t_0+\frac{2\pi q}{p}\right)=1 \, .
\end{equation}
But, given that the set $\{t_0+2\pi q/p\}$ is dense
and that $f(t)$ is a continuous function with
$f(t_0)=1$, it follows that $f(t)$ must be the unit
function.

In order to show that the function $g(t)$ in
(\ref{transform}) vanishes, we go back and consider
the sequences
$\{\sqrt{g_n}\tilde\beta^J_n(t)/(\kappa_n^*)^2\}$, but
specialized now to the case $f(t)=1$. We recall that
the terms of these sequences must tend to zero at
infinite $n$, because of the square summability of
$\{\sqrt{g_n}\tilde\beta^J_n(t)\}$. Then, using again
the asymptotic limits of $\alpha_n(t)$ and
$\beta_n(t)$, we conclude that the sequences given by
\begin{equation}
\label{z2}
\displaystyle
g(t)-4 \frac{\lambda_n}{\kappa^*_n}\omega_n\sin{\left[(n+1)
\tau\right]}e^{-i(n+1)\tau}
\end{equation}
must have a vanishing limit as well. In obtaining this
last expression, we have employed the fact that the
constants $\lambda_n$ must tend to zero when
$n\rightarrow\infty$, as it follows from the unitarity
of the dynamics once $f(t)=1$ has been established
(see \cite{CMV5}).

Finally, it can be proved \cite{CMOV} that the real
and imaginary parts of (\ref{z2}) cannot both tend to
zero for all possible values of $t$ unless the
function $g(t)$ vanishes. As in Appendix A of
\cite{CMV5}, the crucial argument involves Lebesgue
dominated convergence (see e.g. \cite{leb}), which guarantees
that a sequence with terms of the type $\sin^2(nt)$
cannot tend to zero for all values of time in a given
interval.

This concludes the proof that asserts and generalizes
previous results involving uniqueness theorems for
Fock quantizations in non-stationary settings
\cite{unique-gowdy-1,CMV5,CMV8}. In
this case, a unitary implementation of the dynamics
using $SO(4)$-invariant complex structures not only
selects a unique quantum representation of the CCR's,
but also fixes the choice of field, picking up a preferred
dynamics.

\section{Discussion and conclusions}
It is not difficult to extend our result to other
compact spatial manifolds of dimension $d \leq 3$, for
which the analysis of \cite{CMV8} already supports the
uniqueness of the Fock representation for fields
satisfying a Klein-Gordon equation with time-dependent
mass in a(n ``inertially'' foliated) static and
homogenous background. We call
again $\omega_n$ the eigenvalues of the
Laplace-Beltrami operator, forming an increasing
sequence, and $g_n$ the dimension of the corresponding
eigenspace. In addition, we suppose that there exists
a characterization of the complex structures that
are invariant under the symmetries of the field
equations similar to that discussed for $SO(4)$. Then,
one can try and repeat the proof along the lines
explained above for the three-sphere. One can easily
realize that a key point to elucidate whether the function
$g(t)$ may differ from zero is the square summability of the
sequence $\{\sqrt{g_n}/\omega_n\}$. For $d$-spheres, this is the
case only for the circle \cite{CMV5}.
Nonetheless, even in this case, the dynamics of the
new field variables (obtained with $g(t)\neq 0$) can be
implemented unitarily with an invariant complex
structure if and only if the same happens for the
original field variables, so that no new invariant
representation with unitary evolution is permitted by
changing the momentum \cite{CMV5}.

As we have pointed out, a discipline where the proved
uniqueness has deep consequences is in cosmology. Our
result immediately eliminates the ambiguity in the
quantization of fields in such cosmological scenarios,
selecting a preferred quantization and confirming the
robustness of its physical predictions. Moreover, it does so
guaranteeing that unitarity is not lost in the
field dynamics, in spite of the fact that the quantum
field theory is realized in a non-stationary
background.

In particular, this applies to
inflationary cosmological models consisting of a
massive Klein-Gordon scalar field propagating in an FRW
spacetime. In addition to inflation, our
results are relevant for the study of perturbations
around non-stationary homogeneous solutions of the Einstein
equations. For instance, this happens in the case of perfect
fluids in FRW cosmologies (for which the perturbations of the
energy-momentum tensor are isotropic) when the
perturbations are adiabatic as well. It can be shown
that the dynamics of the gauge-invariant energy
density perturbation amplitude, or equivalently of the
Bardeen potential, is dictated then by a wave equation
with a time-varying mass \cite{mukhanov,bardeen}, in a
static and homogeneous spacetime with the spatial
topology of the FRW background. Another important
example is the propagation of gravitational waves in
FRW spacetimes with the topology of a three-sphere,
treated as tensor perturbations \cite{bardeen}. Again,
for isotropic perturbations of the energy-momentum
tensor, these gravitational waves satisfy a field
equation which belongs to the considered class.
Conceptually, there should be no obstruction to apply
our uniqueness theorem to these tensor quantities,
selecting in this way fundamental fields (with
the appropriate dynamics) as well as
preferred Fock representations for them. Finally, we
also mention the perturbations of an FRW universe with
$S^3$-topology whose matter content is a massive
scalar field. These perturbations were studied in
\cite{hall_haw}. The coefficients of the expansion of
the matter perturbations in harmonics, in a suitable
gauge and with an appropriate scaling, satisfy an
equation whose solutions reproduce those for a free
field with a time-dependent mass (in a static
background), up to asymptotic corrections for large
harmonic numbers.

In summary, we have seen that there exists not only
a unique representation for the CCR's that implements
unitarily the dynamics using $SO(4)$-invariant vacuum
states \cite{CMV8}, but we have also reached a deeper
conclusion: there is a unique choice of
fundamental field such that the theory is
compatible with both requests --dynamics and
symmetries. The possible ambiguities of the
quantization are completely removed. In one hand, the
strong results showed here have an immediate
application in physics, particularly in cosmology. On
the other hand, they are a basis to extend powerful
uniqueness theorems in standard quantum mechanics to
quantum field theory.

\section*{Acknowledgments}
We would like to thank R. Jackiw for correspondence. We are also grateful
to the anonymous
referee for suggestions to improve the original version of the manuscript.
This work was supported by the grants CERN/FP/109351/2009 from Portugal,
DGAPA-UNAM IN108309-3 from Mexico,
and MICINN FIS2008-06078-C03-03 and
Consolider-Inge\-nio Program CPAN (CSD2007-00042) from Spain.
J.O. acknowledges CSIC
for the grant No. JAE-Pre\_08\_00791.

\bibliographystyle{plain}

\end{document}